\pdfoutput=1
\documentclass[acmtog,authorversion]{acmart}

\setcitestyle{square}

\acmJournal{TOG}
\acmVolume{37}
\acmNumber{4}
\acmArticle{103}
\acmYear{2018}
\acmMonth{8}

\setcopyright{acmcopyright}

\acmDOI{10.1145/3197517.3201286}

\acmSubmissionID{144}

\usepackage{booktabs}

\usepackage{color}
\definecolor{blue}{rgb}{0,0,1}
\definecolor{red}{rgb}{1,0,0}
\definecolor{green}{rgb}{0,.5,0}
\definecolor{orange}{rgb}{0.75, 0.4, 0}
\newcommand{\rs}[1]{{\color{black}\textbf{}#1}\normalfont}

\newcommand{\hh}[1]{{\color{black}\textbf{}#1}\normalfont}

\newcommand{\etal}{et al.}
\newcommand{\N}{\mathcal{N}}

\begin{document}

\title{Full 3D Reconstruction of Transparent Objects}

\author{Bojian Wu}
\affiliation{%
 \institution{SIAT}
}
\affiliation{%
\institution{Shenzhen University}
}
\affiliation{%
\institution{University of Chinese Academy of Sciences}
}
\author{Yang Zhou}
\affiliation{%
	\institution{Shenzhen University}
}
\affiliation{%
	\institution{Huazhong University of Science \& Technology}
}
\author{Yiming Qian}
\affiliation{%
	\institution{University of Alberta}
}
\author{Minglun Gong}
\affiliation{%
	\institution{Memorial University of Newfoundland}
}
\author{Hui Huang}
\authornote{Corresponding author: Hui Huang (hhzhiyan@gmail.com)}
\affiliation{%
   \department{College of Computer Science \& Software Engineering}
   \institution{Shenzhen University}
}

\renewcommand\shortauthors{B. Wu, Y. Zhou, Y. Qian, M. Gong, and H. Huang}

\begin{abstract}
Numerous techniques have been proposed for reconstructing 3D models for opaque objects in past decades. However, none of them can be directly applied to transparent objects. This paper presents a fully automatic approach for reconstructing complete 3D shapes of transparent objects. Through positioning an object on a turntable, its silhouettes
and light refraction paths under different viewing directions are captured.  Then, starting from an initial rough model generated from space carving, our algorithm progressively optimizes the model under three constraints: surface and refraction normal consistency, surface projection and silhouette consistency, and surface smoothness. Experimental results on both synthetic and real objects demonstrate that our method can successfully recover the complex shapes of transparent objects and faithfully reproduce their light refraction properties.
\end{abstract}


%
%






\maketitle

\section{Introduction}
\label{sec:intro}

Reconstructing 3D models of real objects has been an active research topic in both Computer Vision and Graphics for decades.  A variety of approaches have been proposed for different applications, such as autonomous scanning~\cite{wu:2014:autoscan}, multi-view stereo~\cite{Galliani2015mvstereo}, photometric stereo~\cite{chen07photostereo}, etc.  While these techniques are able to faithfully capture and reconstruct the shapes of opaque or even translucent objects, none of them can be directly applied on transparent objects.  As a result, people often have to paint those  transparent objects before capturing their shapes.

On another front, how transparent objects refract lights toward a fixed viewpoint can be accurately acquired using environment matting techniques~\cite{Chuang:2000:EME,qian:2015:envmat}. Since light refraction paths are determined by surface normals, one has to wonder whether the shape of the transparent object can be inferred accordingly.  This has been demonstrated as a feasible direction in a previous work~\cite{qian:2016:reconstruct}.  Through enforcing a position-normal consistency constraint, their approach can generate point clouds on two sides of a given transparent object. Nevertheless, the captured surface shape is incomplete. In addition, this approach is not that easy to apply, involving setting up and calibrating among 4 different camera/monitor configurations.

\begin{figure}[t!]
	\centering
	\includegraphics[width=\linewidth]{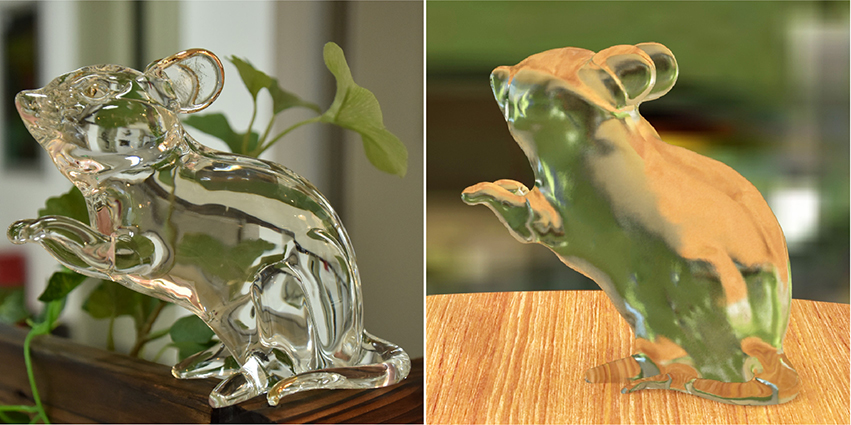}
	\caption{A transparent object refracts lights in its environment (left) and hence its shape cannot be reconstructed using conventional techniques. We present a novel method that, for the first time, directly reconstruct the shapes of transparent objects from captured images. This allows photo-realistic rendering of real transparent objects in virtual environments (right). For a clear view of the reconstructed geometry, please see Fig.~\ref{fig:gallery_bunny_mouse}.
	}
	\label{fig:teaser}
\end{figure}

\rs{This paper presents a fully automatic approach for reconstructing complete 3D shapes of transparent objects with known refractive indexes}. Through positioning the object on a turntable, its silhouettes and light refraction paths under different viewing directions are captured using two fixed cameras (Section~\ref{sec:capture}). An initial model generated through space carving is then gradually evolved toward the accurate object shape using novel point consolidation formulations that are constrained by captured light refraction paths and silhouettes (Section~\ref{sec:recon}). Results on both synthetic and real objects (Section~\ref{sec:results})  demonstrate the effectiveness and robustness of our approach; see e.g., Fig.~\ref{fig:teaser}.

\begin{figure}[t!]
	\centering
	\includegraphics[width=\columnwidth]{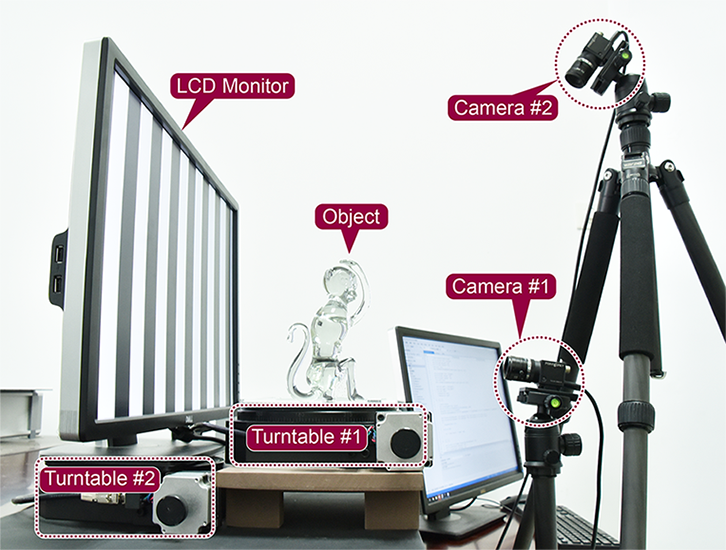}
	\caption{Our data acquisition setup. \rs{The object to be captured (the \emph{Monkey} statue in this case) is placed on Turntable \#1. A LCD monitor is placed on Turntable \#2 and serves as a light source.  Camera \#1 faces the object and the monitor for capturing silhouettes and ray-pixel correspondences. Camera \#2 looks downward to the Turntable \#1 for its rotation axis calibration. The bottom right monitor belongs to a PC that controls the data capture and is not used for illuminating the scene.}}
	\label{fig:setup}
\end{figure}

\section{Related Work}
\label{sec:related}

\paragraph*{Surface reconstruction.}
The literature on 3D surface reconstruction~\cite{Berger2013,Berger2014} is vast.  Specifically, reconstructing transparent objects is well-known a challenging problem~\cite{ihrke2010transparent}. Recent developments, such as reconstruction of flames~\cite{ihrke2004image,wu2015reconstruction}, mixing fluids~\cite{gregson2012stochastic}, gas flow~\cite{atcheson2008time,ji2013reconstructing}, and cloud~\cite{levis2015airborne,levis2017multiple}, aim at dynamic inhomogeneous transparent objects, whereas we focus on static reflective and refractive surfaces with homogeneous materials. Our approach is automatic and non-intrusive, different from existing intrusive acquisition methods~\cite{Aberman:2017:DTS:3072959.3073693,hullin2008fluorescent,trifonov2006tomographic}.

\paragraph*{Environment matting.}
To composite transparent objects into novel backgrounds, environment matting is often applied. This problem is introduced by Zongker \etal~\shortcite{zongker1999environment}, wherein environment mattes are estimated by projecting a series of horizontal and vertical color stripes. Chuang \etal~\shortcite{Chuang:2000:EME} extend the work for locating multiple distinct contributing sources, and also propose a single-image solution for colorless and purely specular objects. Wexler \etal~\shortcite{wexler2002image} develop an image-based method, which allows estimating environment mattes under natural scene background but requires a large amount of sample images. The problem can also be solved in the wavelet~\cite{peers2003wavelet} and frequency~\cite{qian:2015:envmat} domain. Compressive sensing is leveraged to reduce the number of projecting patterns~\cite{duan2015compressive,qian:2015:envmat}.

\paragraph*{Shape-from-X}
To estimate the geometry of transparent objects,  the shape-from-distortion techniques~\cite{Ben-Ezra:2003:MRT:946247.946706,tanaka2016recovering,wetzstein2011refractive} focus on analyzing known or unknown distorted background patterns. Zuo \etal~\shortcite{zuo2015interactive} incorporate internal occluding contours into traditional shape-from-silhouette methods, and propose a visual hull refinement scheme. It is also possible to reconstruct transparent objects by capturing exterior specular highlights~\cite{morris2007reconstructing,yeung2011adequate} known as shape-from-reflectance. However, the acquisition approach requires manually moving a spotlight around the hemisphere to illuminate the object and a reference sphere from different directions. Recent shape-from-polarization methods~\cite{pmvs_CVPR_2017,huynh2010shape,miyazaki2005inverse} connect polarization states of light with shape and surface material properties. Here we utilize the shape-from-silhouette to initialize our reconstruction.

\paragraph*{Direct ray measurements.}
Kutulakos and Steger~\cite{kutulakos2008theory} provide theoretical analysis of the reconstruction feasibility using light path triangulation, and categorize the problem with numbers of reflections or refractions involved. Some researchers only focus on one-refraction events~\cite{shan2012refractive,Yue_caustic,Schwartzburg_caustic}, in particular for fluid surface reconstruction~\cite{morris2011dynamic,Qian_2017_CVPR,zhang2014recovering}. Tsai \etal~\shortcite{tsai2015does} consider two-refraction cases instead. \hh{Note that with given incident and exit ray-ray correspondences~\cite{ji2013reconstructing,wetzstein2011refractive, IseringhausenSIG2017}}, depth-normal ambiguity still exists as they are interrelated with each other. Qian \etal~\shortcite{qian:2016:reconstruct} propose a position-normal consistency constraint for solving the two-refraction reconstruction problem, but they only compute a pair of front-back surface depth maps. Kim \etal~\shortcite{kimacquiring} develop a method to reconstruct axially-symmetric objects that could contain more than two refractions, however, cannot be applied to general non-symmetric objects.

\begin{figure}[t!]
	\centering
	\includegraphics[width=\columnwidth]{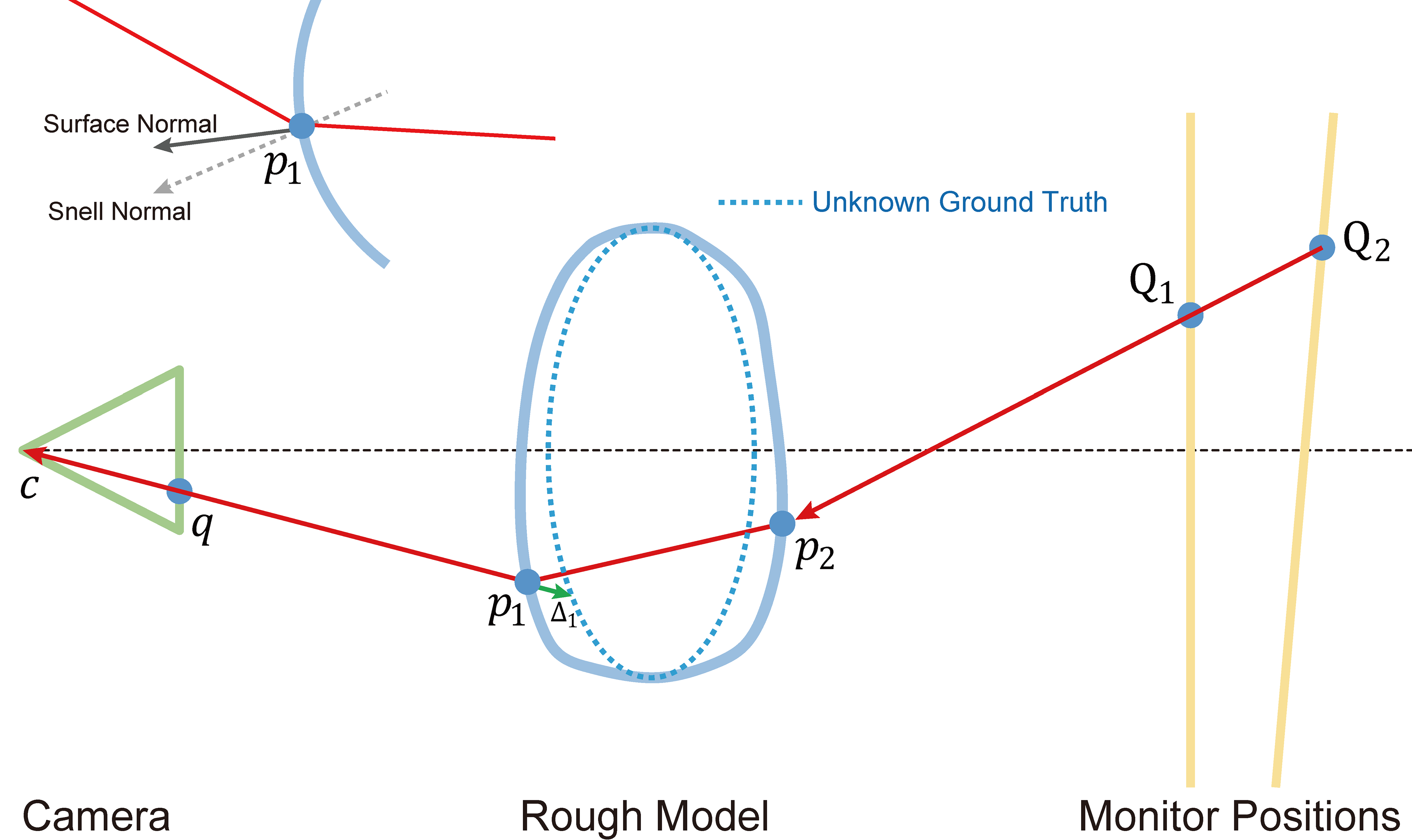}
	\caption{Estimate light refraction path from ray-ray correspondence. For the exit ray that reaches camera center $c$ through pixel $q$, its corresponding incident ray can be acquired by connecting the illuminating pixels $Q_1$ and $Q_2$ on monitor at different positions. Under the assumption that only two refractions happen, a full light path can be formed by linking points $p_1$ and $p_2$ on the rays based on hypothesized surface depths. The true light refraction path is the one that minimizes the differences between the surface normal (estimated from surface shape) and refraction normal (estimated using Snell's law) at both $p_1$ and $p_2$ locations; see inset on the top left corner.}
	\label{fig:ray-ray}
\end{figure}

\begin{figure*}[t!]
	\includegraphics[width=\textwidth]{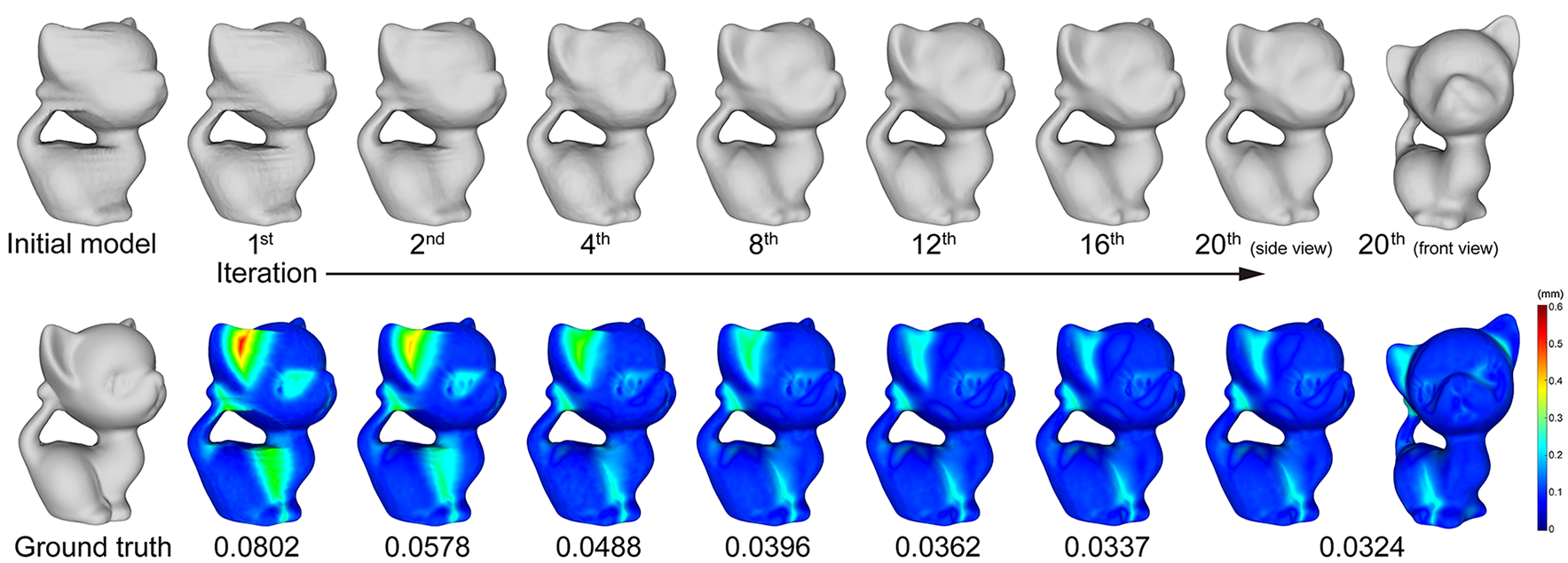}
	\caption{Progressive reconstruction on a synthetic \emph{Kitten} model with refractive index set to 1.15. Top: starting from an initial rough model obtained by space carving, our approach gradually recovers geometric details in concave areas and converges after 20 iterations. \rs{Bottom: the reconstruction error is visualized using the Hausdorff distance (measured per vertex) between the corresponding model and ground truth, forming as an error map. The number below each error map denotes the average Hausdorff distance (in millimeters). Compared to the size of the bounding box ($6.3 \times 9.7 \times 5.7$ mm), the error is quite small.}}
	\label{fig:kitten}
\end{figure*}

\paragraph*{Point data consolidation.} 
Point clouds estimated from the ray-ray acquisition are in general highly unorganized, and corrupted heavily with noise, outliers, overlapping or missing regions~\cite{qian:2016:reconstruct}. A straightforward data cleaning step may easily cause the over-smoothing.  
Inspired by the point projection based data consolidation framework~\cite{huang2009wLOP,EAR2013,lipman2007LOP,Dpoints15},
we define two novel point consolidation formulations that project points sampled from initial rough surfaces toward the latent object geometry based on captured light refraction paths and silhouettes, respectively. Applying the two consolidation formulations in alternating manner can effectively guide the reconstructed model toward the true object shape, whereas directly applying existing data consolidation techniques does not yield satisfiable results.

\section{Capturing setup}
\label{sec:capture}

Here we first explain the setup that we have designed for data acquisition. As shown in Fig.~\ref{fig:setup}, the transparent object to be captured is placed on Turntable \#1. Two cameras are used and both are fixed during the capture process. Camera \#1 is positioned in front of the transparent object and Camera \#2 above it. \rs{Both cameras have their intrinsic parameters and relative positions calibrated~\cite{Zhang_calib}.}  In addition, through putting a checkerboard pattern on the turntable, its rotation axis with respect to the two cameras is also calibrated.

Similar to the previous work~\cite{qian:2016:reconstruct}, a monitor is used as light source. Nonetheless, instead of manually moving the monitor during acquisition to capture starting locations and orientations of incoming rays, we place the monitor on top of Turntable \#2. The monitor's position can then be precisely and automatically adjusted.

To start the acquisition, we use Turntable \#2 to set the monitor at its first position, which is calibrated with the cameras through displaying a checkerboard pattern.  At this monitor position, we rotate the transparent target object using Turntable \#1 to observe it from a set of (8 by default) directions that evenly sample the $360^o$ viewing angle. At each direction, \rs{a series of binary Gray codes are displayed for both silhouette extraction~\cite{zongker1999environment} and environment matting.} The latter allows us to determine the pixel location on monitor that corresponds to a given ray refracted by the object and observed by Camera \#1.

The process is repeated after setting the monitor to its second position. Here the monitor is moved using Turntable \#2, but it can also be moved manually. The object is rotated again to perform environment matting from the exact same set of view directions.  Since the monitor is moved, a new illuminating pixel location can be computed for each observed ray.  Connecting it with the corresponding one obtained in the previous round thus provides the incoming ray orientation; see 2D illustration in Fig.~\ref{fig:ray-ray}.

The images captured using the aforementioned procedure not only provide  ray-ray correspondence, but also allow us to compute object silhouettes at each captured view.  Since extracting the silhouettes requires much less computational effort, a higher sampling density is used to capture additional silhouettes. In practice, 72 view directions that evenly sample the $360^o$ horizontal viewing angle are used for all examples presented in this paper.

\section{Reconstruction Method}
\label{sec:recon}

As described previously, the captured views provide us two important data: 1) silhouettes of the object from different views, which define the visual hull of the object, and 2) ray-ray correspondences before and after rays intersecting with the object, which correlate to light refraction paths and surface geometry details.

Our reconstruction starts from gathering all silhouettes to produce an initial rough model by space carving~\cite{SpaceCarving}. The ultimate goal is to optimize this rough model according to the captured ray-ray correspondences while maintaining shape silhouettes. We achieve it through gradually updating the model under three constraints: surface and refraction normal consistency, surface projection and silhouette consistency, and surface smoothness. Fig.~\ref{fig:kitten} shows our progressive results with reconstruction accuracy measurements on a synthetic \emph{Kitten} example. 
We detail the optimization process in the following three subsections.

\subsection{Surface and refraction normal consistency}
\label{subsec:pnc}
Given a rough model, we first shoot rays from the camera to find the intersections, and then optimize the depths of these intersections along the corresponding rays according to the captured ray-ray correspondences.

As shown in Fig.~\ref{fig:ray-ray}, each ray-ray correspondence captured at a given view associates an exit ray observed by the camera to two pixels on the monitor, one before and one after the monitor is moved. Connecting the two pixel locations gives us the incident ray parameters. All captured ray-ray correspondences are within the object silhouettes and hence intersect with the object. In Qian~\etal's approach~\shortcite{qian:2016:reconstruct}, it is assumed that the light is refracted only twice when traveling between the monitor and the camera. One refraction occurs at the intersection between the exit ray and the object surface (referred to as front intersection), whereas the other occurs at the intersection between the incident rays and the surface (referred to as the back intersection). Under this assumption, directly connecting the front and the back intersections gives us the ray traveling path within the transparent object. The normal needed for achieving the desirable ray refraction effect at each intersection location can also be computed, based on Snell's law~\cite{born2013principles}.

\begin{figure}[t]
	\includegraphics[width=\columnwidth]{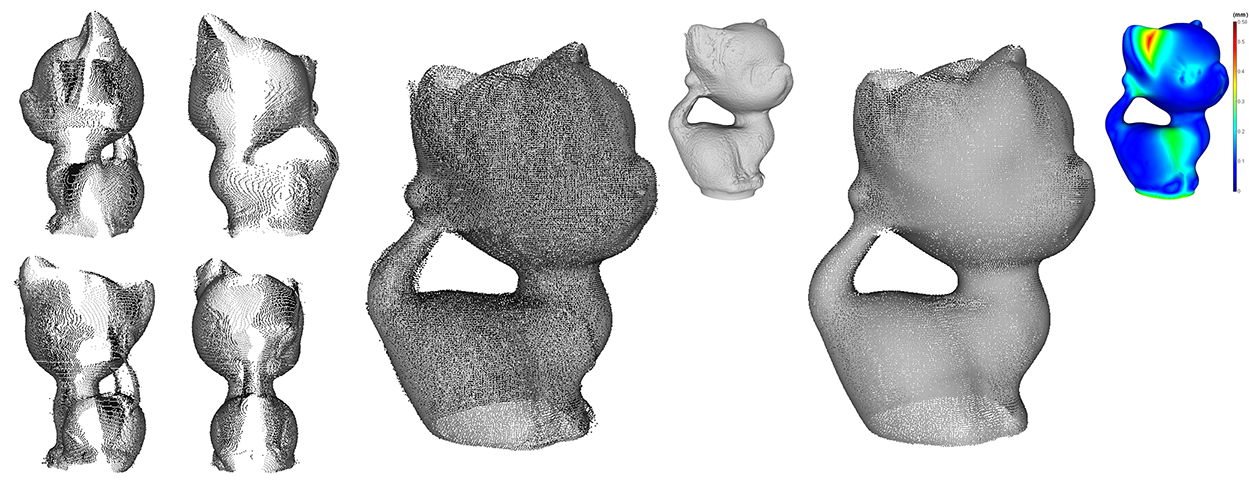}
	(a) \hspace{2.3cm}  (b) \hspace{2.5cm} (c)
	\caption{Model reconstructed using only surface and refraction normal consistency. As in~\cite{qian:2016:reconstruct}, each opposite pair of the eight captured views can be used to generate front and back depth maps, resulting four incomplete point clouds (a). Directly merging them together (b) and feeding into Poisson surface reconstruction (top right in (b)) results in a noisy and inaccurate model. Applying point consolidation~\protect{\cite{huang2009wLOP}} to clean up the data does not fully solve the problem either, as the resulting point cloud (c) is still far away from the ground truth; see the error map color coded on the top right of (c).} 
	\label{fig:nor}
\end{figure}

We adopt the same assumption with certain relaxation. Thanks to our rough model, we are able to trace each individual ray path between the monitor and the camera to filter out the captured ray-ray correspondences that involve more than two intersections. This allows us to reconstruct more complex object shapes (e.g., the \emph{Mouse} statue shown in Fig.~\ref{fig:teaser}), which have more than two refractions under some view directions. \rs{In addition, rays that are involved in total reflections can also be detected and pruned.}

The rough model also provides a good initial solution when optimizing the real surface shape.  Similar to Qian~\etal's approach~\shortcite{qian:2016:reconstruct}, the surface shape is computed implicitly through optimizing the depth of intersection in captured images. That is:
\begin{equation}
\label{eq:pnc}
\min_d { \sum_{i \in I} { \texttt{\big (}  \left\| N(i) - SN(i) \right\|^2 + \lambda \sum_{i'\in\N_i} {\left\| d_i - d_{i'}\right\|^2 }\texttt{\big ) }} },
\end{equation}
where $N(i)$ denotes the surface normal at the intersection point $p_i$, and $SN(i)$ denotes the Snell normal~\cite{qian:2016:reconstruct} induced by Snell's law according to the light path of the refraction happened at the surface point. \rs{We have $d=\left\{d_i~|~i\in I\right\}$ and $d_i$ is the depth from camera to the intersection point $p_i$ along the ray. 
The set $I$ only contains valid intersection points by ray-ray correspondences after removing those of involving more than two refractions and total reflections.
The set $\N_i$ contains the indices of $p_i$'s local neighborhood, which is computed using 4-connected neighboring pixels along the corresponding view.} A standard 2-norm $\left\|\cdot\right\|$ is applied.

The first term minimizes the discrepancy between surface normal (approximated by local PCA analysis) and Snell (refraction) normal~\cite{qian:2016:reconstruct}.
The second term penalizes on depth roughness. \rs{Empirically, we set the balancing parameter $\lambda = 10/\texttt{diaglen}$  by default, where \texttt{diaglen} denotes the diagonal length of the object bounding box. That is, the larger the object is, the smaller the value of $\lambda$ shall be.}
Optimizing both terms produces a depth map for each captured view.  Fig.~\ref{fig:nor} shows the point cloud generated by registering together four views.  The resulting model, even though quite noisy, better captures surface details than the initial rough model, especially in concave regions.

\begin{figure*}[t]
	\includegraphics[width=\linewidth]{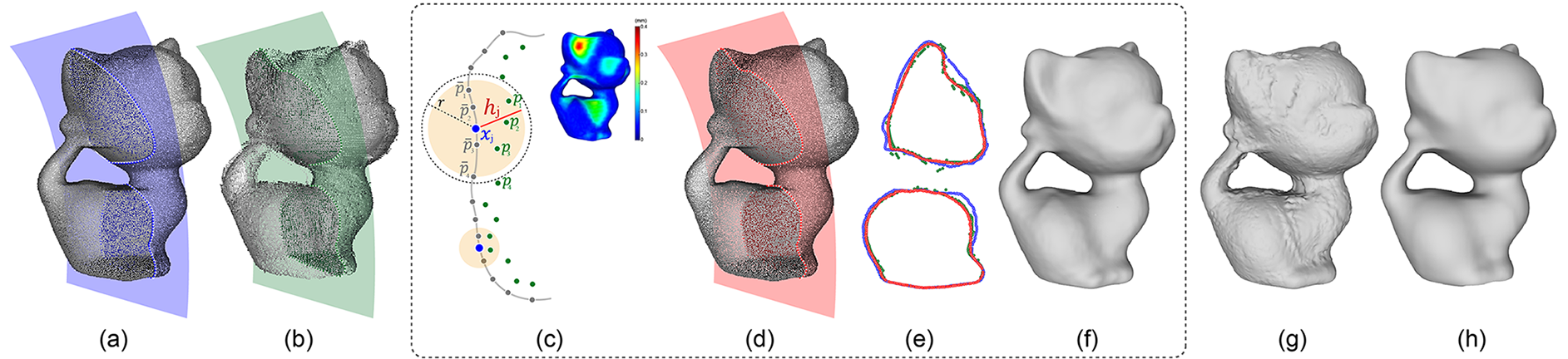}
	\caption{Importance of adaptive neighborhood size selection for point consolidation. For the point set $X$ (in blue) sampled from the initial rough model (a), our approach adaptively selects the neighborhood size $h$ for each point $x$ based on the average distance between the rough model and the noisy point cloud (in green) estimated using ray-ray correspondences (b). The neighborhood sizes computed for different areas are color coded in (c), which shows that larger neighborhoods are used for areas not well modeled by the rough model, such as the concave ear region, and smaller neighborhoods for convex parts. Projecting sample points using adaptive neighborhoods results in a consolidated point cloud (in red) that better captures the shape of the \emph{Kitten} around its ear and back while maintains to be smooth; see cross-section curves shown in (e). Using the consolidated point cloud shown in (d), we can get better reconstructed shape on these concave areas (f). In comparison, projection using a fixed neighborhood size will either leads to a noisy model when the neighborhood size is small (g), or to an over-smoothed model when the neighborhood is large (h). In both cases, the surface in concave regions are not properly reconstructed. }
	\label{fig:AdaProjection}
\end{figure*}

\begin{figure}[t]
	\includegraphics[width=\columnwidth]{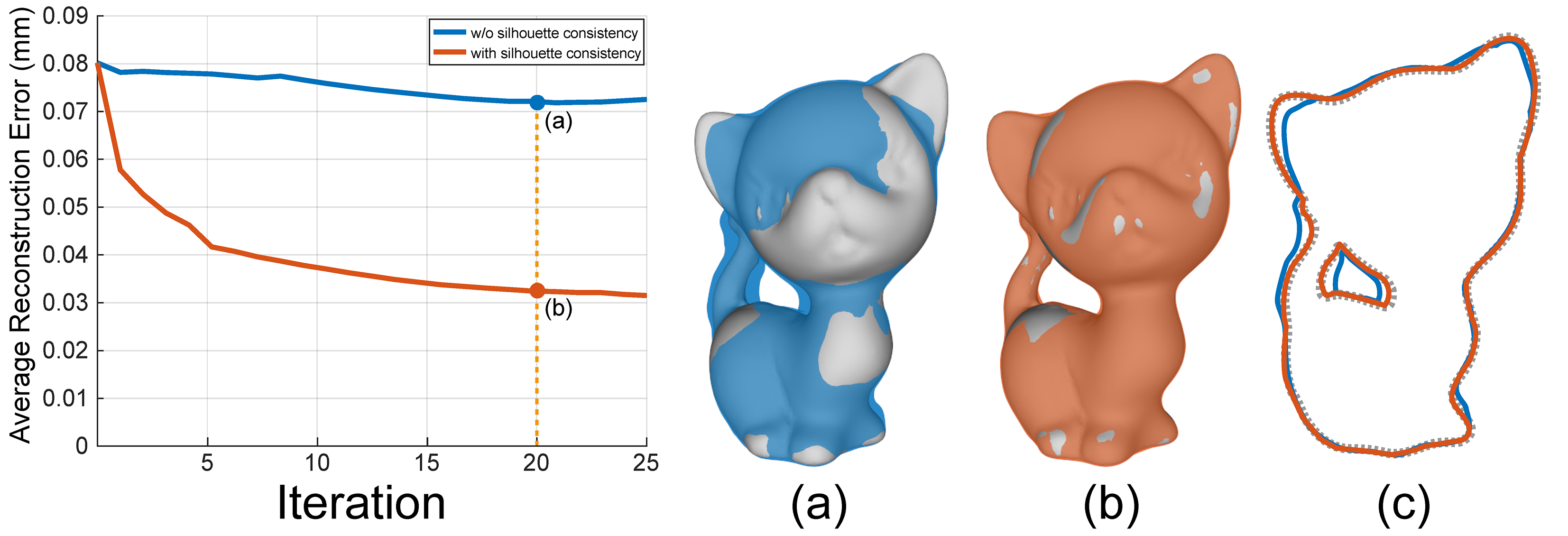}
	\caption{Impact of constraint on shape silhouette consistency. The comparison shows that, when silhouette consistency~\eqref{eq:sil} is applied, the reconstructed mesh (shown in red in (b)) is much closer to the ground truth (shown in gray) in ears, tail and claws regions than the mesh obtained without using this constraint (blue in (a)). The contours of all three shapes are shown in (c) using the same color coding.}
	\label{fig:silhouette}
\end{figure}

\subsection{Surface projection}
\label{subsec:AdaLOP}
As pointed out in Qian~\etal's paper~\shortcite{qian:2016:reconstruct}, the depth maps obtained using surface and refraction normal consistency are often noisy and incomplete.  This is because ray-ray correspondences may not be captured for all pixels within the silhouette and ray refraction is highly sensitive to surface normal. As a result, the point cloud generated from optimizing~\eqref{eq:pnc}, denoted by $P=\left\{ p_i~|~i\in I \right\}$, contains heavy noise, outliers, mis-aligned errors, and missing areas; see e.g., Fig.~\ref{fig:nor} (b).  Not only data consolidation is necessary, directly applying state-of-the-art consolidation techniques without any shape priors does not generate satisfiable results either; see e.g., Fig.~\ref{fig:nor} (c).

To address this challenge, we turn to the initial rough model. \rs{By applying Poisson-disk sampling~\cite{Corsini_sampling} on the rough model, we obtain a set of points (30K by default), $X^0=\left\{x^0_j~|~j\in J\right\}$, that evenly samples this complete, smooth, yet inaccurate surface.} Our strategy is hence to smoothly evolve the point set $X$ to recover more geometric features from the point cloud $P$ while maintaining its completeness and smoothness. 
Inspired by the point consolidation work~\cite{huang2009wLOP,Dpoints15}, given the current iterate $X^k$, $k = 0, 1,...$, \rs{we compute the next iterate $X^{k+1} = \left\{x^{k+1}_j~|~j\in J\right\}$ by minimizing:
\begin{equation}
\sum_{j \in J} \texttt{\Big(}\sum_{i \in I} \| x^{k+1}_j - p_i \| \theta(\| x_j^k - p_i\|) + \frac{\alpha}{|\N_j|} \sum_{j' \in \N_j} \| \Delta_j-\Delta_{j'} \|^2 \texttt{\Big)},
\label{eq:consoli}
\end{equation}
where $\Delta_j = x^{k+1}_j - x_j^k$ is the displacement vector,} and $\theta(\| x_j^k - p_i\|)= e^{-\| x_j^k - p_i\|^2 / (h_j/4)^2 } $ is a fast descending function with an adaptive support radius $h_j$ that defines the size of the influence data neighborhood adaptively with respect to $x_j$. 
We set the weighting parameter $\alpha=7.5$ by default to balance the two terms in~\eqref{eq:consoli}.

Same as previous approaches~\cite{lipman2007LOP,huang2009wLOP}, the first term is a local $\ell_1$-median projection, which is known as an effective noise-removal operator for unorganized point clouds and is non-sensitive to the presence of outliers. However, unlike previous work that uses a regularization term to enforce sample point distribution, here the second term is defined as the the Laplacian on the projection displacements.  Such a change is motivated by the fact that the source of initial samples are different.  In previous approaches~\cite{lipman2007LOP,huang2009wLOP}, initial points are sampled from incomplete point cloud and hence are unevenly distributed.  In our case, the set $X$ is evenly sampled from the initial complete rough surface. Therefore, we only need to maintain the distribution of samples using a simpler Laplacian regularization.

On the other hand, the previous approaches use a fixed neighborhood size for consolidating all points on the surface~\cite{lipman2007LOP,huang2009wLOP}.  In our case, the rough model can be very close to real object shape in areas close to the silhouettes, but dramatically different from it in concave regions.  Hence, using a small neighborhood size cannot effectively project points in concave regions, whereas using a large neighborhood may blur the geometric details we would like to recover in areas near the silhouettes. 

To address this issue, an adaptive neighborhood radius $h_j$ is used for projecting each sample $x_j$.  It is computed based on the average distance between point cloud $P$ and rough model $X$ in the neighborhood of $x_j$. That is, a smaller neighborhood will be used when the point cloud estimated using ray-ray correspondences agrees with the model generated from silhouettes, whereas a larger neighborhood will be used when the two disagree. 
	
In particular, $h_j$ is defined as the average of the ray shooting distances between $p_i$ and the corresponding $\bar{p}_i$ that hits on the rough model and lies in the local neighborhood $\N_{x_j}$:\rs{
\begin{equation}
\label{eq:AdaRadius}
h_j = \sum_{i\in \N_{x_j}}\frac{1}{|\N_{x_j}|}\|p_i - \bar{p}_i\|, \;  \N_{x_j} = \{ i ~|~\|x_j - \bar{p}_i\| \leq r\}.
\end{equation}

\begin{figure}[t!]
	\centering
	\includegraphics[width=\columnwidth]{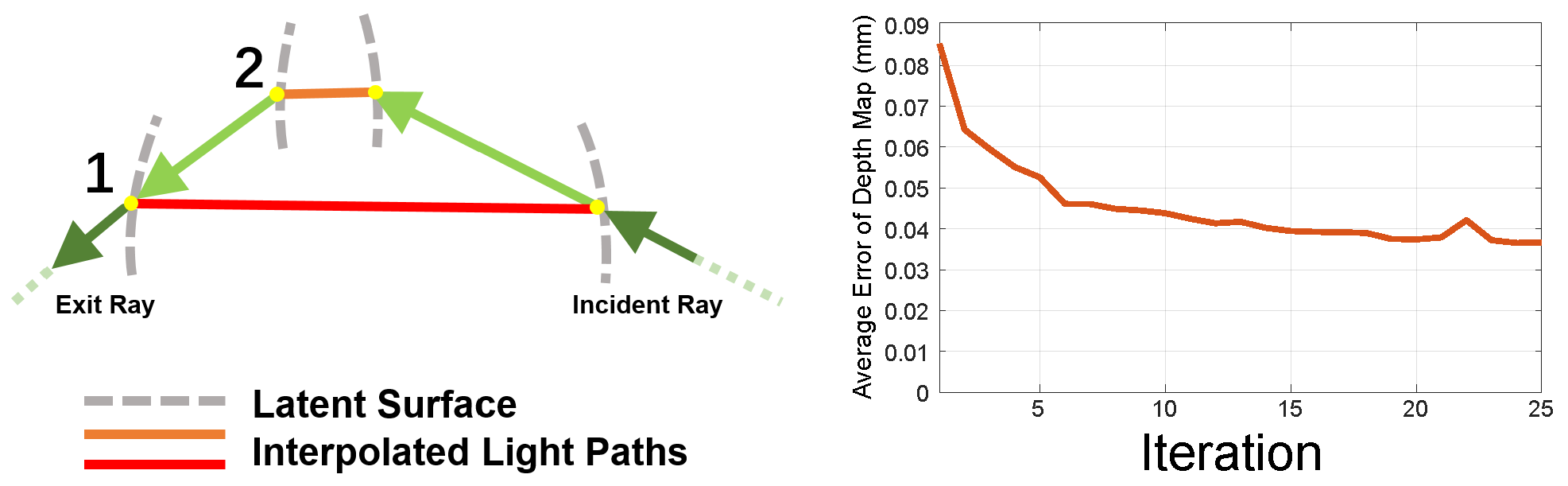}
	\caption{Limitation on normal consistency constraint. For a pair of ray-ray correspondence, there may exist multiple refraction surfaces that satisfy the surface and refraction normal consistency constraint (left). 
		This introduces ambiguities when estimating surface depth using (\protect{\ref{eq:pnc}}). Feeding in better initial solution can alleviate the problem. As a result, with the rough model getting more accurate during the progressive reconstruction, the errors in depth maps obtained by optimizing (\protect{\ref{eq:pnc}}) also get smaller (right).
	}
	\label{fig:WhyProgressive}
\end{figure}

The parameter $r$ is computed when we use Poisson-disk sampling to generate $X^0$ with 30K points by default, i.e., the average distance among initial samples.} Hence, only points $\{\bar{p}_i\}$ whose distance to $x_j$ is less than $r$ are used for computing the average value for $h_j$. The radius $h_j$ is usually large in concave regions and small in areas where the rough model is already approximated well to ground truth. Fig.\ref{fig:AdaProjection} shows the effectiveness of using our adaptive local projection.

\begin{table}[!t]
	\centering
	\caption{\rs{Average computation time per iteration of different processing steps on both synthetic and real examples presented in the paper.} Note that space carving is only performed once for each object and the time for normal consistency is the sum of solving 4 front-back view pairs.}
	\begin{tabular}{lc}
		\hline
		Reconstruction steps   & Time (mins) \\ \hline
		Space carving          & 15          \\
		Normal consistency     & 15          \\
		Surface projection     & 3           \\
		Silhouette consistency & 3           \\
		Screened Poisson       & 0.1         \\ \hline
	\end{tabular}
	\label{tab:time}
\end{table}

\subsection{Silhouette consistency}
\label{subsec:sil}
As discussed above, silhouettes and light refraction paths provide independent cues on the shape of real surface.  Silhouettes offer accurate shape boundary information under selected viewpoints. The light refraction paths provide surface depth cues for both convex and concave areas, but are prone to noises. More importantly, as shown in Fig.~\ref{fig:WhyProgressive}(a), the normal consistency constraint can be ambiguous and hence the estimated surface depth may not be accurate.  Even though the initial rough model obtained through space carving perfectly matches the silhouettes, after applying the aforementioned consolidation step to satisfy the surface and normal consistency, the resulting model may deviate from the captured silhouettes.  It is thus worth to enforce the surface projection and silhouette consistency.

Specifically, we want the projection of the point cloud to fully occupy the captured silhouettes in all views.  This is achieved by minimizing a data term defined using the distance between the boundary of point cloud projection and the captured silhouettes. Combining such a data term with the same smoothing term as defined in~\eqref{eq:consoli} gives us the following objective function:\rs{
\begin{equation}
\label{eq:sil}
\min_{X}\sum_{j \in J}\texttt{\Big(} \sum_{v=1}^V \delta_j^v D(q_j^v,\partial\Omega_v) + \frac{\beta}{|\N_j|} \sum_{j'\in\N_j}\| \Delta_j - \Delta_{j'} \|^2 \texttt{\Big)},
\end{equation}
where $\Delta_j = x_j - x_j^k$ is the displacement vector as in~\eqref{eq:consoli}, and $V$ (defaulted to be 72) is the number of captured silhouettes. We have $q_j^v = \mathbf{P}^v\cdot x_j$ to denote the 2D projection of the sample point on view $v$, where $\mathbf{P}^v$ is the corresponding projection matrix of view $v$.  $\partial\Omega_v$ denotes the boundary of the matting mask $\Omega_v$ on view $v$, i.e., the object silhouette on this view. $\delta_j^v$ is a binary indicator function, which equals to 1 if $q_j^v$ lies on the boundary of the projected shape on view $v$ under a threshold, otherwise becomes 0. The function $D(q_j^v, \partial\Omega)$ returns the closest distance from the point $q_j^v$ to $\partial\Omega$ on the projection image plane. We set the parameter $\beta=50/\texttt{diaglen}$ by default to balance the fitting and smoothness terms with respect to the sizes of transparent objects.}

To compute the binary indicator function $\delta_j^v$ in the data term, the following procedure is used. We start with projecting all sample points $\left\{x_j~|~j \in J\right\}$ onto the given view $v$. The areas covered by the latent surface represented by sample points are determined through a flood fill operation, which is performed on a \emph{k}-NN graph ($k=6$ by default) build upon the 3D points. The boundary contour of the filled 2D shape is then computed and the value of the indicator function $\delta_j^v$ is determined. Please note that, since the projection operation may change the neighborhood structure of the point cloud, the 3D \emph{k}-NN graph and the 2D filled shapes need to be updated after each optimization iteration.

Solving the above silhouette consistency optimization~\eqref{eq:sil} enforces a smooth point-based shape deformation to ensure that the projections of resulting point cloud match well with silhouettes in all captured  views.
Fig.~\ref{fig:silhouette} quantitatively compares the results generated with and without silhouette consistency optimization. The reconstruction error here is measured using the Hausdorff distance between the reconstructed model and the ground truth. The reconstruction error plotted under the two settings provide convincing evidence on the importance of this constraint.

\subsection{Progressive reconstruction}
\begin{figure*}[t!]
	\includegraphics[width=\textwidth]{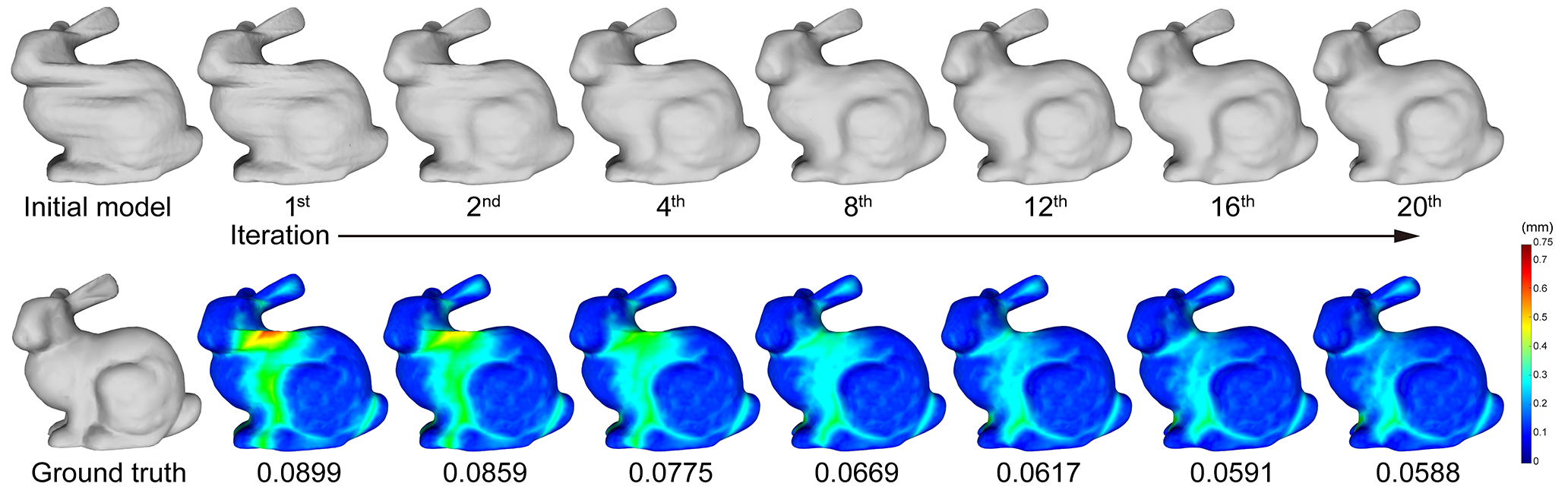}
	\caption{\rs{Results on synthetic \emph{Bunny} with refractive index set to 1.15. The size of the bounding box of \emph{Bunny} is $8.4 \times 8.3 \times 6.5$ mm. The numbers below each error map is the average Hausdorff distance between the reconstructed shape and ground truth.}}
	\label{fig:bunny}
\end{figure*}

Once the point cloud sampled from the initial rough model goes through two phases of consolidation using different constraints, a new 3D surface model is generated from the resulting point cloud using screened Poisson surface reconstruction~\cite{kazhdan2013screened}.  This new model will serve as the rough shape for the next round of sampling, surface depth estimation (Section~\ref{subsec:pnc}), consolidation (Section~\ref{subsec:AdaLOP} and~\ref{subsec:sil}), and reconstruction. \rs{As the rough model gets more accurate, it can more precisely filter out ray-ray correspondences that involve more than two refractions and total reflections and provide better initial solution for optimizing~\eqref{eq:pnc}.} This helps to alleviate the aforementioned ambiguity problem for normal consistency and leads to better surface depth maps; see Fig.~\ref{fig:WhyProgressive}. The overall surface model can therefore progressively approach the true object shape; see Figs.~\ref{fig:kitten} and~\ref{fig:bunny}.

It is worth noting that using two phases of consolidation to satisfy normal consistency and silhouette consistency constraints is necessary and important. In fact, attempts were made to formulate both consistency constraints into a single objective function.  However, since the initial rough model obtained through space carving matches perfectly with all captured silhouettes.  It becomes a local optimal solution and hence, the optimization process is often stalled.  Alternatively consolidating points under the two constraints with the regularization on surface smoothness well balances the stochasticity that is
necessary for searching a global optimal reconstruction.

\section{Results and Discussion}
\label{sec:results}

We have implemented our algorithm in C++, with parallelizable parts optimized using OpenMP. On average, each reconstruction iteration takes about 20 minutes on a 24-core PC with 2.30GHz Xeon CPU and 64GB RAM. \rs{We solve the objective functions \eqref{eq:pnc} and \eqref{eq:sil} by L-BFGS-B~\cite{l-bfgs-b}, and adopt the iterative algorithm proposed in~\cite{huang2009wLOP} to optimize ~\eqref{eq:consoli}. Table~\ref{tab:time} lists the average computation time per iteration for each step of the reconstruction process. As in general our algorithm converges within 20 iterations, the full reconstruction for an object can be completed within 5$\sim$6 hours. In addition, we need roughly another two hours to render or capture all data, and one more hour to compute the alphamatte and ray-ray correspondences for each transparent object.}

\subsection{Synthetic experiments with evaluation}
\begin{figure*}[t!]
	\includegraphics[width=\textwidth]{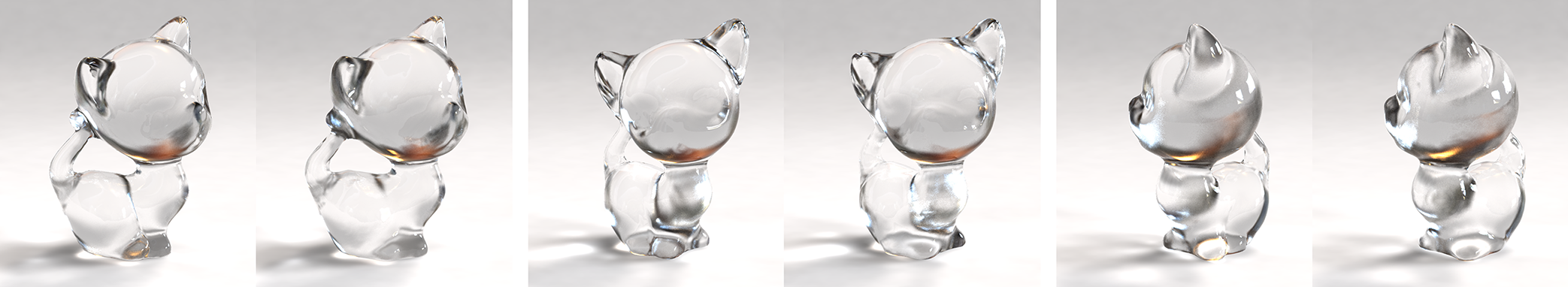}
	\caption{Visual comparison on the rendering results of \emph{Kitten}. Each of the three image pairs is generated using the same view angle, where the left one is the rendering of ground truth and the right is our recovered model.}
	\label{fig:kitten_render}
\end{figure*}

\begin{figure}[t!]
	\includegraphics[width=\columnwidth]{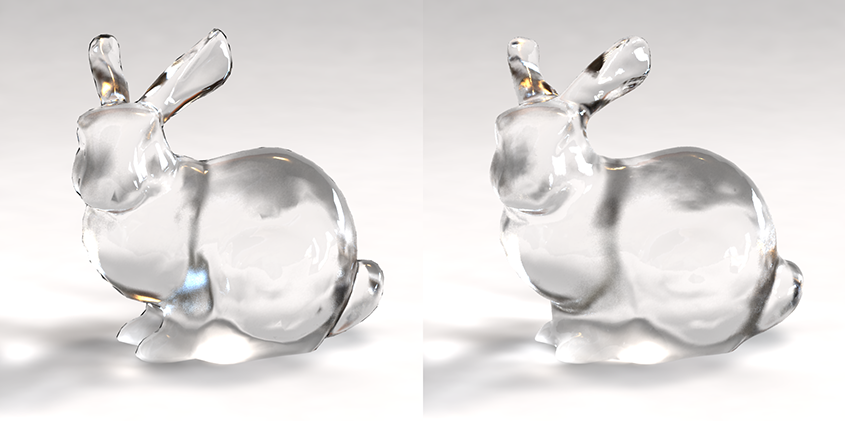}
	\caption{Visual comparison on the rendering results between ground truth \emph{Bunny} (left) and our recovered model (right).}
	\label{fig:bunny_render}
\end{figure}

To evaluate our method, we first run the algorithm on two widely used synthetic models: \emph{Kitten} and \emph{Bunny}. Both models are rendered using POV-Ray\footnote{Persistence of Vision Raytracer, http://www.povray.org/} as transparent objects, with the refractive index set to 1.15. The objects, virtual cameras, and virtual monitors are set up the same way as discussed in Section~\ref{sec:capture}. During the rendering, we turn off anti-aliasing to avoid edge blurriness and to better capture ray-ray correspondences. \rs{Both two virtual cameras are set to be the pinhole model. The resolution of virtual monitor is $1920\times 1080$. Thus, for each view of Camera \#1, it needs to render 22 Gray encoded images, 11 for rows and 11 for columns.

The models progressively reconstructed from these rendered images can be seen from Figs.~\ref{fig:kitten} and~\ref{fig:bunny}. Distances from reconstructed models to the ground truth surface are visualized as error maps for quantitative evaluation.} For both models, the average distance decreases with more iterations, which demonstrates the effectiveness of our algorithm. Figs.~\ref{fig:kitten_render} and~\ref{fig:bunny_render} visually compare the rendering results of ground truth models and our reconstructions in the same environment under different views. They suggest that our approach can nicely reproduce the appearances of transparent objects.

\rs{We also test our algorithm on synthetic examples with different refractive indices in Fig.~\ref{fig:ior}. From the error curve we can see that higher IOR correlates to larger residual error. This is mainly due to the fact that the object is only illuminated from the back. Hence, higher IOR leads to fewer captured ray-ray correspondences, resulting in higher reconstruction error.}

\subsection{Real object experiments with evaluation}
\label{subsec:real}

\begin{figure}[t!]
	\includegraphics[width=\columnwidth]{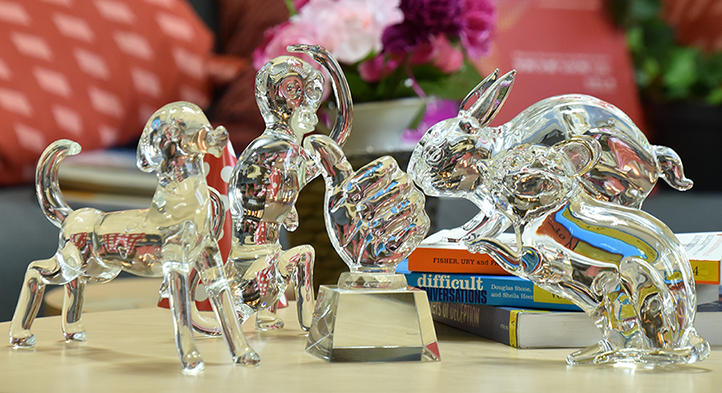}
	\caption{Transparent objects used for testing in this paper. The combination of complex object shapes and cluttered environment leads to high frequency signals in the captured image.}
	\label{fig:real_objects}
\end{figure}

\hh{Five real transparent objects made from borosilicate 3.3 glass\footnote{According to the manufacturer, the glass has a melting point of 820$^\circ$C and a refractive index of 1.4723} are also used for our experiments; see Fig.~\ref{fig:real_objects}. A DELL LCD monitor (U2412M) with resolution $1920\times 1200$ is used for displaying Gray code patterns, where 22 images (11 for rows and 11 for columns) are needed under each view of Camera \#1. During capturing, the brightness and contrast of this monitor are set to be the highest for sufficient background illumination. The turntable is controlled by a $1.8^{\circ}$ 57mm stepper motor with a gear ratio of 180:1, possessing repeatability accuracy of $0.005^{\circ}$. To capture data for real objects, two Point Grey Flea3 color cameras (FL3-U3-13S2C-CS) are used. While Camera \#2 uses the default settings with its focus on Turntable \#1, we change Camera \#1 to be the manual mode and set a small aperture (about f/6.0) in order to mimic the pinhole model. Also, Camera \#1 is focused on the object for clear imaging. To optimize the brightness and quality of captured images, the shutter time and gain of the camera are set to 40$\sim$50ms and 0dB, respectively.}

\begin{figure}[t!]
	\includegraphics[width=\columnwidth]{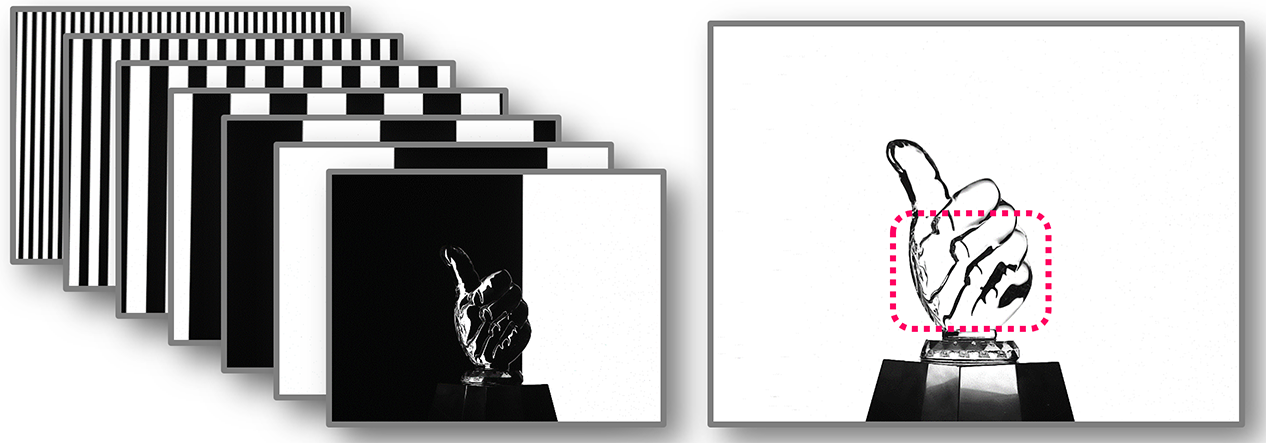}
	(a) \hspace{4cm}  (b) 
	\caption{Examples of captured images of a \emph{Hand} object(a). However, as highlighted in (b), even when all pixels on the monitor are turned on, there are still areas on the object do not get illuminated, as they refract lights from directions not covered by the background monitor. Their ray-ray correspondences, therefore, cannot be captured.} 
	\label{fig:hand_captured}
\end{figure}

Fig.~\ref{fig:hand_captured} shows the images captured for the \emph{Hand} object under one of the views. Using ray-ray correspondences extracted from these images, 
we are able to recover surface depths under each view through optimizing surface and refraction normal consistency; see Fig.~\ref{fig:hand}. Directly consolidating these depth information does not provide a satisfiable model. Our algorithm, on the other hand, starts from a rough model obtained using space carving and progressively enriches it as shown in Fig.~\ref{fig:hand_evolve} (a). The final converged model is smooth and nicely captures surface details in concave areas.

To conduct quantitative evaluation on this real object, we reconstructed its ground truth shape using an intrusive method. As shown in Figs.~\ref{fig:hand_evolve}(b-d), we painted it with DPT-5 developer, and scanned it with a high-end industrial level scanner. The reconstruction errors of models generated by our approach are then evaluated through registering them with the ground truth model using ICP and then computing the average distance in-between them. As shown in Fig.~\ref{fig:hand_evolve}(e), when dealing with real objects, our progressive reconstruction approach converges equally well as with synthetic data.  Even though there is still residual reconstruction error in the end, our result improves significantly (by 26 percent) over the initial model obtained by space carving. 

\begin{figure*}[t!]
	\includegraphics[width=\textwidth]{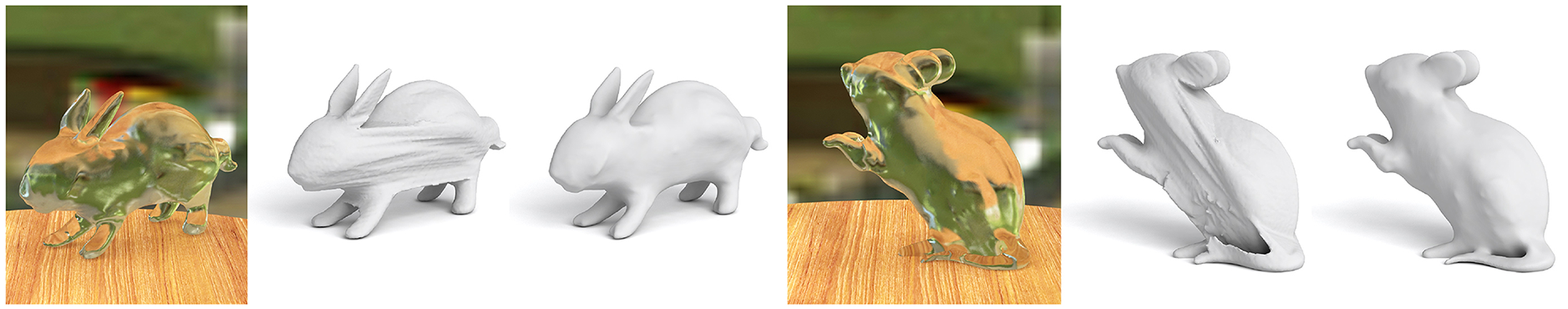}
	\caption{Reconstructions of the real transparent \emph{Bunny} and \emph{Mouse}. In each image group, the left image shows the photo-realistic rendering of our recovered model, whereas the middle and right images are used to compare the initial rough model with the final reconstruction. Note how the neck and tummy regions of both shapes are significantly improved, while the silhouettes of the object are well preserved.}
	\label{fig:gallery_bunny_mouse}
\end{figure*}

\begin{figure*}[t!]
	\includegraphics[width=\textwidth]{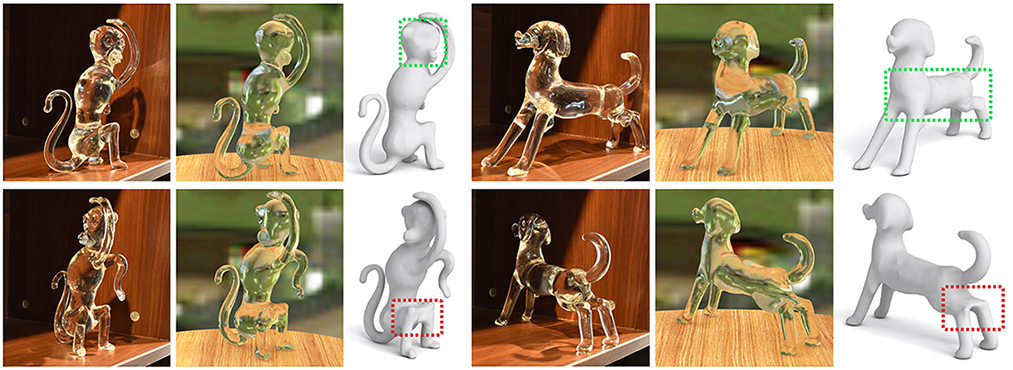}
	\caption{Reconstruction results for \emph{Monkey} and \emph{Dog} under two view directions. For each object, the reconstruction (right column) successfully captures the featured concave parts (highlighted in green box). Nonetheless, we can still observe areas not well-reconstructed (highlighted in red) due to the violation of two-refraction assumption. Since these reconstruction artifacts only show up in areas involving multiple refractions, they are hardly noticeable when comparing the rendering of our reconstructed model (middle column) with real object photos (left column).}
	\label{fig:gallery_monkey_dog}
\end{figure*}

Fig.~\ref{fig:gallery_bunny_mouse} shows reconstruction results on the transparent \emph{Bunny} and \emph{Mouse} objects. Our final reconstruction provides noticeable improvement on the concave areas (e.g., neck and tummy) over the initial rough model, while the silhouette contours (e.g., leg and back) of the object are well preserved. Fig.~\ref{fig:gallery_monkey_dog} shows another two results on the transparent \emph{Monkey} and \emph{Dog}. Our reconstruction successfully recover the eye region for the monkey and the belly shape for the dog. However, the crotch areas of both two models are not well-recovered due to the violation of two-refraction assumption.

\subsection{Discussions on reconstruction error}
\begin{figure}[t!]
	\includegraphics[width=\columnwidth]{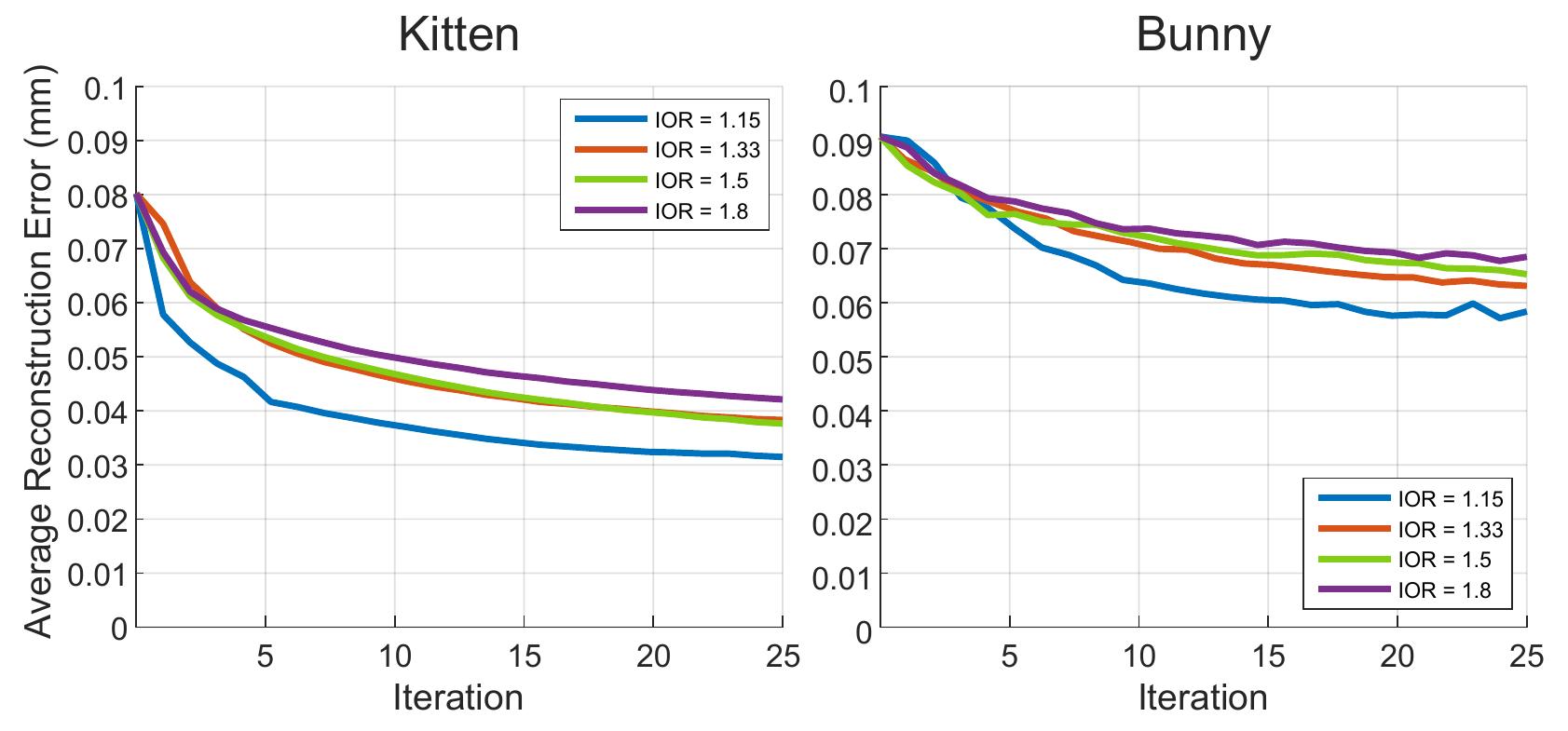}
	\caption{\rs{Synthetic tests with different refractive indexes. The average reconstruction error after each iteration is plotted for \emph{Kitten} and \emph{Bunny}.}}
	\label{fig:ior}
\end{figure}

\begin{figure*}[t!]
	\includegraphics[width=\textwidth]{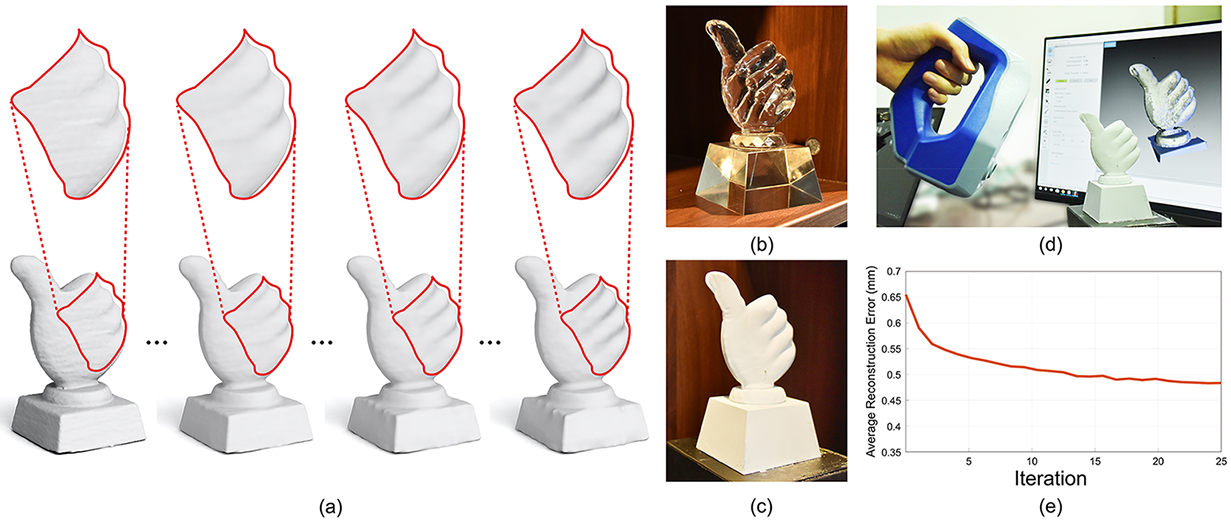}
	\caption{Progressive reconstruction of the \emph{Hand} object shown in Fig.~\ref{fig:hand_captured}. Our approach gradually recovers surface details that are not available in the initial rough model (a). To conduct quantitative evaluation, we also painted the transparent object (b) with DPT-5 developer (see (c)) and then carefully scanned it using a high-end Artec Space Spider scanner (d). \rs{The size of the bounding box of this scanned \emph{Hand} model is $80 \times 119 \times 64$ mm.} Using this captured model as ground truth, the average reconstruction error after each iteration is plotted in (e). The result curve shows that our approach can effectively reduce reconstruction errors and converge after 20 iterations. }
	\label{fig:hand_evolve}
\end{figure*}

\hh{As shown in Figs.~\ref{fig:kitten}, \ref{fig:bunny}, \ref{fig:ior} and~\ref{fig:hand_evolve}, the residual reconstruction errors are less than $1/100$ of the object size.  Based on our setup in synthetic experiments, each pixel roughly projects to 0.015mm in length on the object's surface. For real data, each pixel from Camera \#1 corresponds to 0.18mm on the object. Correspondingly, we can convert the average residual error into about 2 pixels for \emph{Kitten} and \emph{Bunny} synthetic data and 3 pixels for the real \emph{Hand} model. 

Our analysis suggests two main sources for the residual error. First, as mentioned in Section~\ref{subsec:sil}, whether the projected point is on the silhouette boundary is determined at pixel level. Thus, the recovered model after optimization with silhouette consistency might deviate from the ground truth for up to a pixel. The other main error source should be derived from the uncertainties of ray-ray correspondences. Practically, in capturing stage, besides the precision issue of ray-pixel extraction on captured images, the sensitivity of refraction and complexity of the surface geometry could introduce many unreliable correspondences, which would directly affect the reliability of the captured ray-ray correspondences. As Fig.~\ref{fig:ior} suggests, the more unreliable ray-ray correspondences are, the higher average reconstruction error would be. For the real \emph{Hand} model in Fig.~\ref{fig:hand_evolve}, due to the finite thickness of DPT-5 developer layer and/or the possible misalignment during the scanning, extra errors might be introduced into the measurement of reconstruction accuracy.}

\begin{figure}[t!]
	\includegraphics[width=\columnwidth]{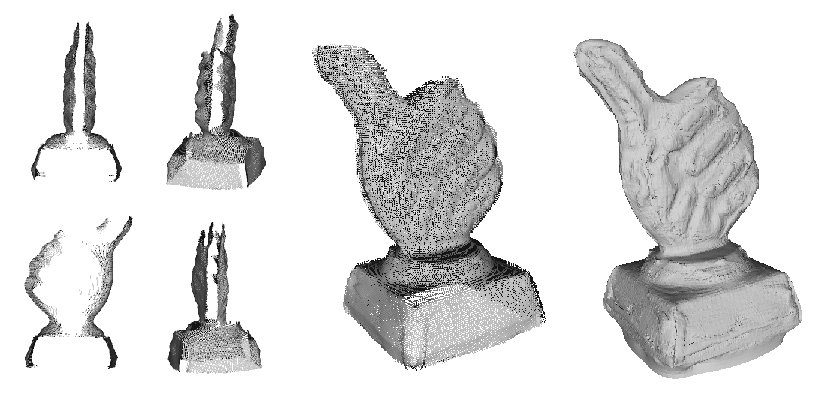}
	\caption{Using data captured from each front-back view pairs, we can estimate front and back surface depths based on surface and refraction normal consistency, resulting a set of incomplete point clouds as shown in left. Directly merging these point clouds (middle) cannot produce satisfactory surface reconstruction (right).}
	\label{fig:hand}
\end{figure}

It is worth noting that, even though our approach still has room for improvement in terms of reconstruction accuracy, it is the first non-intrusive and fully-automatic approach for reconstructing complete 3D shapes of transparent objects. In comparison, Qian \etal~\shortcite{qian:2016:reconstruct}'s method only recovers the incomplete point clouds for front and back surfaces. As shown in Figs.~\ref{fig:nor} and~\ref{fig:hand}, directly merging the incomplete point clouds under different views does not provide satisfactory results. To recover full 3D models for transparent objects, we formulated a novel silhouette constraint and used it to gradually optimize the reconstructed shape by iterating between the normal consistency and silhouette constraint. Since capturing silhouettes is much easier and more reliable than ray-ray correspondences, our reconstruction is more reliable and robust than Qian \etal~\shortcite{qian:2016:reconstruct}.

\section{Conclusions and Future Work}
\label{sec:conclusion}

This paper presents the first practical method for automatically and directly reconstructing complete 3D models for transparent objects based only on their appearances in a controlled environment. The environment is designed using affordable and off-the-shelf products, which include a LCD monitor, two turntables, and two cameras. This setup can work in fully automatic fashion, removing the needs for manually adjusting object positions and calibrating the cameras.

Two sets of data are captured, one is for shape silhouettes and the other is for ray-ray correspondences before and after light refraction. Our presented algorithm fully utilizes both sets of data and progressively reconstructs the 3D model of a given transparent object using three constraints: surface and refraction normal consistency, surface projection and silhouette consistency, and surface smoothness.  Experiments on both synthetic and real objects with quantitative evaluations demonstrate the effectiveness of our algorithm.

Our method still has several limitations, which set up our future work. The first one is about the data capturing process. As shown in Fig.~\ref{fig:hand_captured}, with one LCD monitor serving as light source behind the object, and one single camera in the front, not all ray-ray correspondences information can be captured, resulting missing data in estimated point cloud. If there is an area that is missing in all captured views, its surface can only be inferred based on surface smoothness and hence geometry details can be lost.
This limitation can be addressed through adding either additional monitors or additional cameras to cover ray-ray correspondence paths with more oblique angles, but at higher system and computation cost.

Secondly, our approach inherited the assumption from~\cite{qian:2016:reconstruct} that the transparent object is homogeneous and only two refractions occur on each light path.  Even though our algorithm can automatically filter out data that violates the two-refraction assumption, and can reconstruct objects that refract lights more than twice in certain directions, it cannot handle multiple refractions directly. This limitation leads to the reconstruction artifacts shown in Fig.~\ref{fig:gallery_monkey_dog}.  In addition, transparent objects that are hollow inside cannot be processed since light is refracted 4 times in all directions.  
\rs{
Also, here we assume the refractive index of the transparent object is known or can be estimated in advance. 
In the near future, we would like to extend our approach to address this full 3D reconstruction problem of transparent objects without these assumptions.}

\section*{Acknowledgments}
We thank the anonymous reviewers for their valuable comments. This work was supported in part by NSFC (61522213, 61761146002, 6171101466), 973 Program (2015CB352501), Guangdong Science and Technology Program (2015A030312015), Shenzhen Innovation Program (KQJSCX20170727101233642, JCYJ20151015151249564) and NSERC (293127). 

\bibliographystyle{ACM-Reference-Format}
\bibliography{FRT}

\end{document}